\documentstyle[12pt, epsf]{article}

\catcode`\@=11
\long\def\@makefntext#1{
\protect\noindent \hbox to 3.2pt {\hskip-.9pt  
$^{{\ninerm\@thefnmark}}$\hfil}#1\hfill}                

\def\@makefnmark{\hbox to 0pt{$^{\@thefnmark}$\hss}}  
        
\def\ps@myheadings{\let\@mkboth\@gobbletwo
\def\@oddhead{\hbox{}
\rightmark\hfil\ninerm\thepage}   
\def\@oddfoot{}\def\@evenhead{\ninerm\thepage\hfil
\leftmark\hbox{}}\def\@evenfoot{}
\def\sectionmark##1{}\def\subsectionmark##1{}}

\renewcommand{\thefootnote}{\fnsymbol{footnote}}

\newcounter{sectionc}\newcounter{subsectionc}\newcounter{subsubsectionc}
\renewcommand{\section}[1] {\vspace*{0.6cm}\addtocounter{sectionc}{1} 
\setcounter{subsectionc}{0}\setcounter{subsubsectionc}{0}\noindent 
        {\normalsize\bf\thesectionc. #1}\par\vspace*{0.4cm}}
\renewcommand{\subsection}[1] {\vspace*{0.6cm}\addtocounter{subsectionc}{1} 
        \setcounter{subsubsectionc}{0}\noindent 
        {\normalsize\it\thesectionc.\thesubsectionc. #1}\par\vspace*{0.4cm}}
\renewcommand{\subsubsection}[1]
{\vspace*{0.6cm}\addtocounter{subsubsectionc}{1}
        \noindent {\normalsize\rm\thesectionc.\thesubsectionc.\thesubsubsectionc. 
        #1}\par\vspace*{0.4cm}}

\newcounter{appendixc}
\newcounter{subappendixc}[appendixc]
\newcounter{subsubappendixc}[subappendixc]

\renewcommand{\appendix}[1] {\vspace*{0.6cm}
        \refstepcounter{appendixc}
        \setcounter{figure}{0}
        \setcounter{table}{0}
        \setcounter{equation}{0}
        \renewcommand{\thefigure}{\Alph{appendixc}.\arabic{figure}}
        \renewcommand{\thetable}{\Alph{appendixc}.\arabic{table}}
        \renewcommand{\theappendixc}{\Alph{appendixc}}
        \renewcommand{\theequation}{\Alph{appendixc}.\arabic{equation}}
        \noindent{\bf Appendix \theappendixc #1}\par\vspace*{0.4cm}}

\def\abstracts#1{{
        \centering{\begin{minipage}{12.2truecm}\footnotesize\baselineskip=12pt\noindent
        \centerline{\footnotesize ABSTRACT}\vspace*{0.3cm}
        \parindent=0pt #1
        \end{minipage}}\par}} 

\renewenvironment{thebibliography}[1]
        {\begin{list}{\arabic{enumi}.}
        {\usecounter{enumi}\setlength{\parsep}{0pt}
\setlength{\leftmargin 1.25cm}{\rightmargin 0pt}
         \setlength{\itemsep}{0pt} \settowidth
        {\labelwidth}{#1.}\sloppy}}{\end{list}}

\newcounter{itemlistc}
\newcounter{romanlistc}
\newcounter{alphlistc}
\newcounter{arabiclistc}

\newcommand{\fcaption}[1]{
        \refstepcounter{figure}
        \setbox\@tempboxa = \hbox{\footnotesize Fig.~\thefigure. #1}
        \ifdim \wd\@tempboxa > 6in
           {\begin{center}
        \parbox{6in}{\footnotesize\baselineskip=12pt Fig.~\thefigure. #1}
            \end{center}}
        \else
             {\begin{center}
             {\footnotesize Fig.~\thefigure. #1}
              \end{center}}
        \fi}

\newcommand{\tcaption}[1]{
        \refstepcounter{table}
        \setbox\@tempboxa = \hbox{\footnotesize Table~\thetable. #1}
        \ifdim \wd\@tempboxa > 6in
           {\begin{center}
        \parbox{6in}{\footnotesize\baselineskip=12pt Table~\thetable. #1}
            \end{center}}
        \else
             {\begin{center}
             {\footnotesize Table~\thetable. #1}
              \end{center}}
        \fi}

\def\@citex[#1]#2{\if@filesw\immediate\write\@auxout
        {\string\citation{#2}}\fi
\def\@citea{}\@cite{\@for\@citeb:=#2\do
        {\@citea\def\@citea{,}\@ifundefined
        {b@\@citeb}{{\bf ?}\@warning
        {Citation `\@citeb' on page \thepage \space undefined}}
        {\csname b@\@citeb\endcsname}}}{#1}}

\newif\if@cghi
\def\cite{\@cghitrue\@ifnextchar [{\@tempswatrue
        \@citex}{\@tempswafalse\@citex[]}}
\def\citelow{\@cghifalse\@ifnextchar [{\@tempswatrue
        \@citex}{\@tempswafalse\@citex[]}}
\def\@cite#1#2{{$\null^{#1}$\if@tempswa\typeout
        {IJCGA warning: optional citation argument 
        ignored: `#2'} \fi}}

 1
 1
 1

\font\ninerm=cmr9

\newcommand\beq{\begin{equation}}
\newcommand\eeq{\end{equation}}
\newcommand\st{\sin\theta}
\newcommand\ct{\cos\theta}
\newcommand\esc{{e \over{\st \ct}}}
\newcommand\half{{1\over{2}}}
\newcommand{\NPB}[1]{{\it Nucl. Phys.}\ {\bf B{#1}}}
\newcommand{\PLB}[1]{{\it Phys. Lett.}\ {\bf B{#1}}}
\newcommand{\PRD}[1]{{\it Phys. Rev.}\ {\bf D{#1}}}
\newcommand{\PRL}[1]{{\it Phys. Rev. Lett.}\ {\bf #1}}

\newcommand{\hc}{ {\rm h.c.} }
\newcommand{\ME}{ M_{ETC} }
\newcommand{\gE}{ g_{ETC} }

\newcommand{\gae}{$\stackrel{>}{\sim}$}

\begin{document}

\centerline{\normalsize\bf TESTING EXTENDED TECHNICOLOR WITH $R_b$}
\baselineskip=16pt
\centerline{\normalsize\bf AND SINGLE TOP-QUARK
  PRODUCTION\footnote{Talk given at the Ringberg Workshop: The Higgs
  Puzzle -- What Can We Learn from LEP2, LHC, NLC and FMC?, Schloss
  Ringberg, Germany, 8-13 December 1996.}}

\vspace*{0.6cm}
\centerline{\footnotesize ELIZABETH H. SIMMONS}
\baselineskip=16pt
\centerline{\footnotesize\it Physics Department, Boston University,
590 Commonwealth Ave.}
\baselineskip=12pt
\centerline{\footnotesize\it Boston, MA,  02215, USA}
\centerline{\footnotesize E-mail: simmons@bu.edu}

\bigskip
\begin{picture}(0,0)(0,0)
\put(295,200){BUHEP-97-5}
\put(290,185){hep-ph/9702261}
\end{picture}
\vspace{24pt}

\vspace*{0.9cm}
\abstracts{We review the connection between $m_t$ and
the $Zb\bar b$ vertex in ETC models and discuss how data on $R_b$
constrains ETC models.  Theories in which the ETC and weak gauge groups do
not commute are consistent with electroweak data and predict effects on
single top production that will be visible at Fermilab.}
 
\normalsize\baselineskip=15pt
\setcounter{footnote}{0}
\renewcommand{\thefootnote}{\alph{footnote}}
\section{Introduction}
Two outstanding questions in particle theory are the cause of
electroweak symmetry breaking and the origin of the masses and mixings
of the fermions.  Because theories that use light, weakly-coupled
scalar bosons to answer these questions suffer from the hierarchy and 
triviality problems, it is interesting to consider the possibility that
electroweak symmetry breaking arises from strong dynamics at scales of
order 1~TeV.  This talk focuses on extended\cite{ETC}
technicolor\cite{tc} (ETC) models,
in which both the masses of the weak gauge bosons and those of the
fermions arise from gauge dynamics.

In extended technicolor models, the large mass of the top quark
generally arises from ETC dynamics at relatively
low energy scales.  Since the magnitude of the
CKM matrix element $\vert V_{tb}\vert$ is nearly
unity, $SU(2)_W$ gauge invariance insures that ETC bosons coupling to
the left-handed top quark couple with equal strength to the
left-handed bottom quark.   In particular, the ETC dynamics
which generate the top quark's mass also couple to the left-handed
bottom quark thereby affecting the $Zb\bar b$ and $Wtb$ vertices\cite{zbbone}.
  
This talk discusses how measurements of $R_b$ constrain ETC model
building, shows that models in which $SU(2)_W$ is embedded in the ETC
group are consistent with experimental data, and explains how 
measurements of single top quark production at the Fermilab Tevatron's Run 3
will further test ETC.

\vfill\eject
\section{From $m_t$ To A Signal of ETC Dynamics}

%
\begin{figure}
\epsfxsize=2.8truein
\centerline{\epsffile{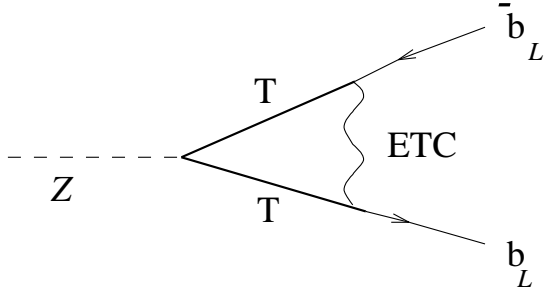}}
\caption{Direct correction to the $Zb\bar b$ vertex
from exchange of the ETC gauge boson that gives rise to the top quark
mass.  Technifermions are denoted by `T'.}
\label{figehs:1}
\end{figure}

Consider an ETC model in which $m_t$ is generated by the exchange of  a
weak-singlet ETC gauge boson of mass $M_{ETC}$ coupling  with
strength $\gE$ to the current
\begin{equation}
{\xi} {\bar\psi^i}_L \gamma^\mu T_L^{ik}
+ {1\over\xi} {\bar t_R} \gamma^\mu U_R^k\ ,\ \ \ \ \ \ {\rm where}\ \ 
\psi_L\ \equiv\ \pmatrix{t \cr b \cr}_L\ \ 
T_L\ \equiv\ \pmatrix{U \cr D \cr}_L 
\label{tmasscur}
\end{equation}
where $U$ and $D$ are technifermions, $i$ and $k$ are weak and technicolor
indices, and $\xi$ is an ETC Clebsch expected
to be of order one.  At energies below $\ME$, ETC  gauge boson exchange
may be approximated by local four-fermion operators.   For example, $m_t$
arises from an operator coupling the  left- and right-handed currents
in Eq. (\ref{tmasscur}) 
\beq
   - {\gE^2 \over  \ME^2}  \left({\bar\psi}_L^i \gamma^\mu
T_L^{iw}\right) \left( {\bar U^w}_R \gamma_\mu t_R \right) + \hc\ 
\label{topff}
\eeq
Assuming, for simplicity, that there is only one weak doublet of
technifermions and that technicolor respects an $SU(2)_L \times
SU(2)_R$ chiral symmetry (so that the technipion decay constant, $F$,
is $v= 246$ GeV) the rules of naive dimensional analysis\cite{dimanal}
give an estimate of
\beq
   m_t\ = {\gE^2 \over \ME^2}
   \langle{\bar U}U\rangle\ \approx\ {\gE^2 \over \ME^2} (4\pi v^3)\ .
\label{topmass}
\eeq
for the top quark mass when the technifermions' chiral
symmetries break.

The ETC boson responsible for producing $m_t$ also affects the $Zb\bar
b$ vertex\cite{zbbone} when exchanged between the two left-handed
fermion currents of Eq. (\ref{tmasscur}) as in Fig. \ref{figehs:1} (with
$T \equiv D_L$ since the ETC boson is a
weak singlet).
This diagram alters 
the $Z$-boson's tree-level coupling to left-handed bottom quarks $g_L =
\esc(-\half + {1\over 3}\st^2)$ by\cite{zbbone}
\beq
\delta g_L^{ETC}\ = \ -{\xi^2 \over 2} {\gE^2 v^2\over\ME^2} \esc(I_3)
\ =\ {1\over 4} {\xi^2} {m_t\over{4\pi v}}
\cdot \esc \label{tb}
\eeq
where the right-most expression follows from applying eq. (\ref{topmass}).

To show that $\delta g_L$ provides a test of ETC dynamics, we must
relate it to a shift in the value of an experimental observable.
Because ETC gives a direct correction to the $Zb\bar b$ vertex, we
need an observable that is particularly sensitive to direct, rather
than oblique\cite{ST}, effects.  A natural choice is the ratio of $Z$
decay widths
\beq
R_b \equiv {\Gamma(Z\to b\bar b) \over {\Gamma(Z \to {\rm hadrons})}}
\eeq
since both the oblique and QCD corrections largely cancel in this ratio.
One finds
\beq
{\delta R_b \over R_b} \approx -5.1\% \xi^2 \left({m_t
\over 175{\rm GeV}} \right).
\label{rbb}
\eeq

Such a large shift in $R_b$ would be readily detectable in current
electroweak data.  In fact, the experimental\cite{LEPdata} value of
$R_b = 0.2179\pm 0.0012$ lies close enough to the standard model
prediction\cite{langacker} (.2158) that a 5\% reduction in
$R_b$ is excluded at better than the 10$\sigma$ level.   
ETC models in which the ETC and weak gauge groups
commute are therefore ruled out.

\section{Non-commuting ETC Models}

The next logical step is to examine models in which the weak and ETC
gauge groups do not commute. 
We begin by describing the symmetry-breaking pattern that enables
``non-commuting'' ETC models to include both a heavy top quark and
approximate Cabibbo universality\cite{NCETC}.  A heavy top quark must
receive its 
mass from ETC dynamics at low energy scales; if the ETC bosons
responsible for $m_t$ are weak-charged, the weak group $SU(2)_{heavy}$
under which $(t,b)_L$ is a doublet must be embedded in the low-scale
ETC group.  Conversely, the light quarks and leptons cannot be charged
under the low-scale ETC group lest they also receive large
contributions to their masses; hence the weak $SU(2)_{light}$ group
for the light quarks and leptons must be distinct from
$SU(2)_{heavy}$.  To approximately preserve low-energy Cabibbo
universality the two weak $SU(2)$'s must break to their diagonal
subgroup before technicolor dynamically breaks the remaining
electroweak symmetry.  The resulting
symmetry-breaking pattern is:

\begin{eqnarray}
ETC & \times& SU(2)_{light} \times U(1)' \nonumber\\ 
&\downarrow&\ \ \ \ \ f \nonumber \\
TC \times SU(2)_{heavy} & \times& SU(2)_{light} \times U(1)_Y  \nonumber\\ 
&\downarrow&\ \ \ \ \ u \\ 
TC & \times& SU(2)_W \times U(1)_Y \nonumber \\
&\downarrow&\ \ \ \ \ v \nonumber \\
TC & \times& U(1)_{EM}, \nonumber  
\end{eqnarray}

\noindent{where $ETC$ and $TC$ stand, respectively, for the extended
technicolor and technicolor gauge groups, while $f$, $u$, and 
$v = 246$ GeV are the expectation values of the order parameters for the
three different symmetry breakings.  Note that, since we are
interested in the physics associated with top-quark mass generation,
only $t_L$, $b_L$ and $t_R$ {\bf must} transform non-trivially under $ETC$.
However, to ensure anomaly cancelation we take both $(t,b)_L$ and
$(\nu_\tau,\tau)$ to be 
doublets under $SU(2)_{heavy}$ but singlets under $SU(2)_{light}$,
while all other left-handed ordinary fermions have the opposite
$SU(2)$ assignment.}

Once again, the dynamics responsible for generating the top quark's
mass contributes to $R_b$.  This time the ETC gauge boson involved
transforms as a weak doublet coupling to
\beq
\xi\bar\psi_L \gamma^\mu U_L +
{1\over \xi}\bar t_R \gamma^\mu T_R
\eeq
where $\psi_L \equiv (t,b)_L$ and $T_R \equiv (U,D)_R$, are doublets
under $SU(2)_{heavy}$ while $U_L$ is an $SU(2)_{heavy}$ singlet.
The one-loop diagram involving exchange of this boson (Figure 1 with $T
\equiv U_L$)
shifts the coupling of  $b_L$ to the $Z$ boson by
\beq
\delta g_L = -\esc {\xi^2 v^2 \over { 2 f^2}} \approx - 
{\xi^2 \over 4} \esc {m_t \over {4 \pi v}} .
\label{a:1}
\eeq
 Since the tree-level $Z b_L
\bar b_L$ coupling is also negative, the ETC-induced change tends to
{\bf increase} the coupling.  Hence $R_b$ increases by\cite{NCETC}
\beq
{\delta R_b \over R_b} \approx +5.1\% \xi^2 \left({m_t
\over 175{\rm GeV}} \right).
\eeq
The change is similar in size to what was obtained in the
commuting ETC models (Eq. (6)), but is opposite in sign.

But that is not the full story of $R_b$ in non-commuting ETC.  Recall
that there are two sets of weak gauge bosons which mix at the scale
$u$.  Of the resulting mass eigenstates, one set is heavy and couples
mainly to the third-generation fermions while the other set is {\it
nearly} identical to the $W$ and $Z$ of the standard model.  That
`nearly' is important: it leads to a shift in the light $Z$'s coupling
to the $b$ of order\cite{NCETC}
\beq
\delta g_L = {e \over{2 \st \ct}} {g_{ETC}^2 v^2\over u^2} \sin^2\alpha
\eeq
where $\tan\alpha = g_{light}/g_{heavy}$ is the ratio of the $SU(2)$
gauge couplings.  The couplings of the light $Z$ to other fermions are
similarly affected.  Mixing thus alters
$R_b$ by
\beq
{\delta R_b \over R_b} \approx -5.1\%  
\sin^2\alpha {f^2\over u^2} \left({m_t
 \over 175{\rm GeV}} \right).
\eeq
Because the two effects on $R_b$ in non-commuting ETC models
are of similar size and opposite sign, these theories can
yield values of $R_b$ that are consistent with experiment\cite{NCETC}.

Since $R_b$ alone cannot confirm or exclude non-commuting ETC, we
should apply a broader set of precision electroweak tests.  This
requires describing the $SU(2)\times SU(2)$ symmetry breaking
sector in more detail.  The two simplest possibilities for the
$SU(2)_{heavy} \times SU(2)_{light}$ transformation properties of the
order parameters that mix and
break these gauge groups are:
\begin{eqnarray}
&\langle \varphi \rangle& \sim (2,1)_{1/2},\ \ \ \ \langle
\sigma\rangle \sim (2,2)_0 ~,\ \ \ \ \ \ \ ``{\rm heavy\ case}"\\
&\langle \varphi \rangle& \sim (1,2)_{1/2},\ \ \ \ \langle
\sigma\rangle \sim (2,2)_0 ~,\ \ \ \ \ \ \ ``{\rm light\ case}"~.
\end{eqnarray}
Here the order parameter $\langle\varphi\rangle$ is responsible for
breaking $SU(2)_L$ while $\langle\sigma\rangle$ mixes
$SU(2)_{heavy}$ with $SU(2)_{light}$.  We refer to these two
possibilities as ``heavy'' and ``light'' according to whether 
$\langle\varphi\rangle$
transforms non-trivially under $SU(2)_{heavy}$ or $SU(2)_{light}$.
In the heavy case, the technifermion condensation responsible for
providing mass
for the third generation of quarks and leptons is also responsible for
the bulk of electroweak symmetry breaking.  The light case corresponds to the
opposite scenario in which different physics provides mass to the
third generation fermions and the weak gauge bosons. 

We have performed\cite{NCETC} a global fit for the parameters of the
non-commuting ETC model ($s^2$, $1/x \equiv v^2/u^2$, and the $\delta
g$'s) to all precision electroweak data: the $Z$ line shape, forward
backward asymmetries, $\tau$ polarization, and left-right asymmetry
measured at LEP and SLC; the $W$ mass measured at FNAL and UA2; the
electron and neutrino neutral current couplings determined by
deep-inelastic scattering; the degree of atomic parity violation
measured in Cesium; and the ratio of the decay widths of $\tau \to \mu
\nu \bar\nu$ and $\mu\to e \nu \bar \nu$.  We find that both the heavy
and light cases provide a good fit to the data.  Furthermore, the extra
$W$ and $Z$ bosons can be relatively light\footnote{These mass limits are
  stronger than current limits from direct searches\cite{dirs} for heavy weak
bosons at FNAL.}.  Figure 2 displays the 95\%
confidence level lower bound (heavy solid line) on the heavy $W$ mass
($M^H_{W}$) for different values of $s^2$ (with $\alpha_s(M_Z)=0.115$);
at large $s^2$, the extra $W$ can weigh as little as 400 GeV.
In the heavy case, similar work shows that the lowest possible heavy $W$
mass at the 95\% confidence level is $\approx 1.6$ TeV, for $0.7< s^2
<0.8$.

We conclude that non-commuting ETC is consistent with all electroweak
tests proposed so far.  Clearly a new test is needed!  In the last
section of the talk, we show that single top quark production may fit
the bill.

\section{Single Top Production}

It has been suggested\cite{wilstel} that a sensitive measurement of the
$Wtb$ coupling can be made at the Tevatron collider by studying single
top production through quark/anti-quark annihilation\cite{stp} $(q\bar
q'\to W \to t b)$, and normalizing to the Drell-Yan process $(q\bar q'
\to W q\to \ell\nu)$ to control theoretical systematic uncertainties
(e.g.  in the initial parton distributions).  This method should be more
precise than alternative methods involving single top production via
$W$-gluon fusion\cite{wgf}, because there is no similar way to eliminate
the uncertainty associated with the gluon distribution function.

In the standard model, the ratio of single top production and Drell-Yan
cross-sections
\begin{equation} {\sigma(q\bar q' \to W \to t b) \over  {\sigma(q\bar
      q'\to  W \to \ell\nu)}} \equiv R^{SM}_\sigma 
\label{oneptone}
\end{equation} 
is proportional to the top quark decay width $\Gamma(t\to W b)$ and,
therefore, to $\vert V_{tb}\vert^2$.  Recent work\cite{heinson} has
shown that with a 30 fb$^{-1}$ data sample from Run 3 at the Tevatron
with $\sqrt{s} = 2$ TeV it should be possible to use single top-quark
production to measure $R_\sigma$, and hence $\vert
V_{tb} \vert^2$ in the standard model, to an accuracy of at least
$\pm$8\%.  By that time, the theoretical accuracy in the standard
model calculation is projected to become at least this good\cite{wilsm}.

The enlarged gauge group in non-commuting ETC models provides two
potential sources of non-standard contributions to $R_\sigma$.
Exchange of ETC gauge bosons can potentially make a large direct
correction to the $Wtb$ vertex, similar to the direct effect on the
$Zb\bar b$ vertex. Furthermore, these models include two
sets of $W$ bosons; both sets contribute to the
cross-sections, and mixing between the two sets alters the couplings of
the lighter $W$ state to fermions.  If the resulting fractional change in the
cross-section ratio $\Delta R_\sigma / R_\sigma$
is at least 16\%, it should be detectable in Run 3.

\begin{figure}[htb]
\epsfxsize 10cm \centerline{\epsffile{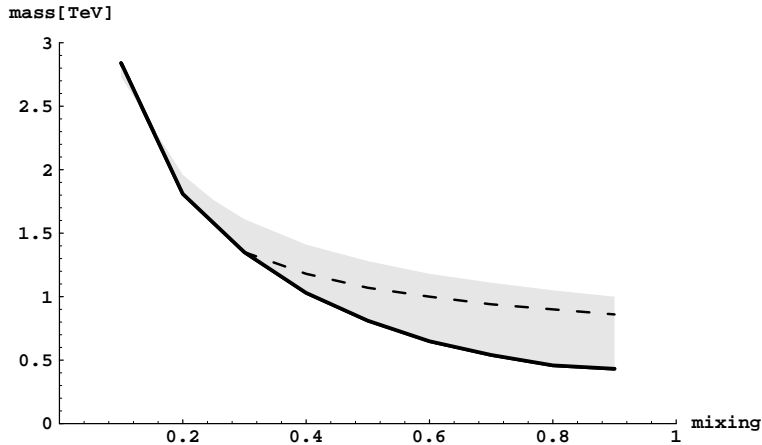}}
\caption[ncetclim]{Region (shaded) where light-case non-commuting ETC models
  predict a visible increase ($\Delta R_\sigma/ R_\sigma \geq 16$\%) in
  single top quark production at TeV33.  The dark line marks the lower
  bound (at 95\% c.l.) on the mass of the heavy weak bosons $M_{W^H}$ (as a function of
  mixing parameter $\sin^2\phi$) by electroweak data\cite{NCETC}.
  Below the dashed line, the predicted value of $\Delta R_\sigma/
  R_\sigma \geq 24$\% .}
\label{ncetclim}
\end{figure}

{\it A priori}, it appears that the $Wtb$ vertex should be affected by
ETC gauge boson exchange through a diagram similar to Figure 1.
However, a closer look at the operator that gives rise to the top quark
mass (the product of the currents in Eq. (8)) demonstrates that there
are no direct ETC contributions to the $Wtb$ vertex of order $m_t/4\pi
v$ in non-commuting ETC models. Because the left-left piece of this
operator includes $(t_l, b_l, U_L)$ but not $D_L$ and because its purely
right-handed piece contains $(t_R,U_R,D_R)$ but not $b_R$, this operator
does {\bf not} contribute to the $Wtb$ vertex.

On the other hand, the presence of two sets of weak boson does alter 
$R_\sigma$.  Diagonalizing the mass matrix of the $W$ bosons yields the
masses, widths, and couplings to fermions of the $W$ mass eigenstates.
Using this information, we have calculated\cite{wtbpaper} the size of
$\Delta R_\sigma/ R_\sigma$ in both the heavy and light cases of
non-commuting ETC.  In the heavy case, the constraint $M_{W^H}$ \gae 1.6
TeV from electroweak data prevents $\vert\Delta R_\sigma/ R_\sigma\vert$
from exceeding 9\%.  This effect is too small to be clearly visible at
Tev33.

The light case of non-commuting ETC, where $M_{W^H}$ can be as
small as 400 GeV, yields more encouraging results.  Since lighter
extra $W$ bosons produce larger shifts in $R_\sigma$, there is a significant
overlap between the experimentally allowed portion of parameter space
and the region in which $\vert \Delta R_\sigma/ R_\sigma\vert \geq
16$\%, as shown in Figure 2.  In fact, the predicted fractional shift in
$R_\sigma$ is greater than 24\% for much of this overlap region.  More
precisely, the shift in $R_\sigma$ is towards values exceeding
$R_\sigma^{SM}$, so that non-commuting ETC models with the ``light''
symmetry breaking pattern predict a visible {\bf increase} in the rate
of single top-quark production.

What allows the corrections to single top-quark production to be
relatively large in non-commuting ETC models is the fact that there is
no direct ETC effect on the $Wtb$ vertex to cancel the contributions
from weak gauge boson mixing.  This is in contrast to the calculation of
$R_b$, where such a cancelation does occur.  Hence within the context
of these models it is possible for $R_b$ to have a value close to the
standard model prediction while $R_\sigma$ is visibly altered.

Finally, we note that\cite{wtbpaper} no model other than light non-commuting ETC has
been found to predict a visible increase in $R_\sigma$.  Thus, single top
quark production can provide a clear signal of dynamical electroweak
symmetry breaking.

\section{Conclusions}

Extended technicolor models predict distinctive alterations in the
$Zb\bar b$ coupling and the rate of single top quark production.
Measurements of $R_b$ have already excluded models in which the ETC and
weak gauge groups commute, in favor of ``non-commuting'' models.
Studies of single top quark production in Run 3 at the Tevatron will
provide the next stringent test of non-commuting ETC.

\medskip
This work was supported in part by NSF grants PHY-9057173 and
PHY-9501249, and by DOE grant
DE-FG02-91ER40676.

\end{document}